\documentclass[iop]{emulateapj}

\usepackage{amsmath, natbib}
\usepackage[usenames,dvipsnames]{color}
\usepackage[colorlinks=true, linkcolor=BrickRed, citecolor=blue, urlcolor=blue, filecolor=blue]{hyperref}
\usepackage[normalem]{ulem}

\setcitestyle{authoryear}

\def\fesc{ f_\mathrm{esc} }
\def\trec{ \bar{t}_\mathrm{rec} }
\def\nH{ \bar{n}_\mathrm{H} }
\def\QHII{ Q_\mathrm{HII} }
\def\CHII{ C_\mathrm{HII} }
\def\MUV{ M_\mathrm{UV} }
\def\LUV{ L_\mathrm{UV} }
\def\MSF{ M_\mathrm{SF} }
\def\xiion{ \xi_\mathrm{ion} }
\def\ngamma{ \dot n_\gamma }
\def\zreion{ z_\mathrm{reion} }
\def\vev#1{ \left\langle #1 \right \rangle }
\def\|{\, | \,}

\def\eg{\emph{e.g.},}

\providecommand{\e}[1]{\ensuremath{\times 10^{#1}}}

\shorttitle{Reconstructing ionizing flux during reionization}
\shortauthors{L.C. Price, H. Trac, \& R. Cen}

\begin{document}

\title{Reconstructing the redshift evolution of escaped ionizing flux from early galaxies with Planck and HST observations}

\author{Layne C. Price,\altaffilmark{1}, Hy Trac,\altaffilmark{1} \& Renyue Cen\altaffilmark{2}}
\affil{\altaffilmark{1}McWilliams Center for Cosmology, Department of Physics, Carnegie Mellon University, Pittsburgh, PA 15213, USA \\
\altaffilmark{2}Department of Astrophysical Sciences, Princeton University, Princeton, NJ 08544, USA}
\email{laynep@andrew.cmu.edu}
\email{hytrac@andrew.cmu.edu}
\email{cen@astro.princeton.edu}

\begin{abstract}

While galaxies at $6 \lesssim z \lesssim 10$ are believed to dominate the epoch of cosmic reionization, the escape fraction of ionizing flux $f_\mathrm{esc}$ and the photon production rate $\dot n_\gamma$ from these galaxies must vary with redshift to simultaneously match CMB and low-redshift observations.  We constrain $f_\mathrm{esc}(z)$ and $\dot n_\gamma(z)$ with \emph{Planck} 2015 measurements of the Thomson optical depth $\tau$, recent low multipole E-mode polarization measurements from \emph{Planck} 2016, SDSS BAO data, and $3 \lesssim z \lesssim 10$ galaxy observations.  We compare different galaxy luminosity functions that are calibrated to HST observations, using both parametric and non-parametric statistical methods that marginalize over the effective clumping factor $C_\mathrm{HII}$, the LyC production efficiency $\xi_\mathrm{ion}$, and the time-evolution of the UV limiting magnitude $dM_\mathrm{SF}/dz$.  Using a power-law model, we find $f_\mathrm{esc} \lesssim 0.5$ at $z=8$ with slope $\beta \gtrsim 2.0$ at $68\%$ confidence with little dependence on the galaxy luminosity function or data, although there is non-negligible probability for no redshift evolution $\beta \sim 0$ or small escape fraction $f_\mathrm{esc} \sim 10^{-2}$.  A non-parametric form for $f_\mathrm{esc}(z)$ evolves significantly with redshift, yielding $f_\mathrm{esc} \sim 0.2, 0.3, 0.6$ at $z=6,9,12$, respectively.  However, a model-independent reconstruction of $\dot n_\gamma(z)$ predicts a suppressed escaped photon production rate at $z=9$ for the latest \emph{Planck} data compared to the other models, implying a quicker period of reionization.  We find evidence for redshift evolution in the limiting magnitude of the galaxy luminosity function for empirical models of the galaxy luminosity function.

\end{abstract}

\section{Introduction}

During the Epoch of Reionization (EoR) the neutral intergalactic medium (IGM) underwent a phase transition into an ionized state.  This transition was mediated by luminous sources that formed shortly after the Big Bang, with overdense regions generating a significant flux of ionizing photons that eventually filtered out into underdense regions of the universe.
Measurements of the Gunn-Peterson optical depth in high redshift quasars~\citep[\eg][]{Becker:2001ee,Fan:2005es,McGreer:2011dm,McGreer:2014qwa} imply that reionization was largely completed by $\zreion \sim 6$.  Similarly, CMB measurements of the Thomson optical depth are sensitive to the mean redshift of reionization, with different combinations of CMB and large-scale structure data predicting a wide range of possibilities for $\zreion$ and the physics of reionization.

Early galaxies are a natural explanation to provide the necessary ionizing photons at redshift $6 \lesssim z \lesssim 12$ (\emph{e.g.}, see recent studies by \citet{Kuhlen:2012vy,Bouwens:2015vha}).
Recent observations of the slope of the galaxy luminosity function provide more evidence that low-luminosity dwarf galaxies can produce enough ionizing photons for reionization~\citep[\eg][]{Bouwens:2010gp,McLure:2012fk,Bouwens:2014fua,finkelstein2015evolution,Livermore:2016mbs}.
While other intriguing sources of ionizing flux might contribute, such as Population III~\citep[\eg][]{2003ApJ...591...12C,2003ApJ...588L..69W,2007ApJ...659..890W} stars and active galactic nuclei~\citep[\eg][]{Madau:1998cd,2003ApJ...586..693W,Madau:2015cga}, in this paper we focus on galaxy-driven hydrogen reionization.

The redshift at which reionization is completed, as well as the duration of this epoch, depends on the density $\ngamma$ of Lyman-continuum (LyC) photons emitted into the IGM in a given unit of time and the hydrogen recombination time $\trec$.  In turn, these depend on a number of relatively uncertain parameters, including the shape and redshift evolution of the galaxy luminosity function (GLF) $\phi$, the time-dependent escape fraction of LyC photons per luminous source $\fesc$, the faint-end magnitude limit $\MSF$ for $\phi$, the effective clumping factor in ionized hydrogen $\CHII$, and the LyC photon production efficiency $\xiion$.  While high-resolution numerical simulations can reduce the theoretical uncertainty in the presumed value of $\CHII$ or $\phi(z)$,
the required complexity makes it challenging to obtain realistic expectations of all of these quantities together and how they evolve with $z$.
It is therefore paramount that we are able to provide a self-consistent, data-based constraint on $\ngamma(z)$ and $\fesc(z)$, including their joint dependencies on the other parameters, in a model-indepent fashion.

The escaped photon production rate density $\ngamma$ is the parameter that can be constrained most effectively from the Thomson optical depth.  Importantly, it can be estimated independently of $\phi(z)$, $\xiion$, or $\fesc(z)$.
However, constraints on $\fesc(z)$ inform us about galactic physics during the reionization epoch, assuming a given GLF with a low-luminosity limiting magnitude $\MSF \sim -10.0$.
Values of $\fesc \gtrsim 0.10-0.20$ are typically required for galaxy-driven reionization models at $z \gtrsim 6$~\citep{Bolton:2007fw}, which conflicts with the low-redshift measured values of $\fesc \lesssim 0.05-0.1$~\citep{2007ApJ...667L.125C,Iwata:2008qx,2016arXiv160201555S}.  This generically requires significant redshift evolution in the escape fraction~\citep[\eg][]{Inoue:2006um,Ferrara:2012zy,Khaire:2015pna}.
However, different numerical models predict a range of functional forms for $\fesc$.  These simulations have not yet arrived at a consensus regarding its expected magnitude or its dependencies on redshift, halo mass, or galaxy mass; for example, it is not yet fully understood if the escape fraction increases or decreases with increasing halo mass~\citep[\eg][]{Wise:2008cs,Razoumov:2009rd,Yajima:2010ij}.
Furthermore, the predicted escape fraction can also vary as a function of metallicity or stellar binary interactions~\citep{Stanway:2015ord,Ma:2016znf}, allowing measurements of $\fesc(z)$ to influence high redshift stellar population models.

\citet{Razoumov:2009rd} find large values of $\fesc \sim 0.8$ at $z=10$ and \citet{Sharma:2015utv} finds $\fesc \gtrsim 0.05-0.20$ at $z >6$ in the \textsc{Eagle} simulation~\citep{Schaye:2014tpa,Crain:2015poa}.  Similarly, \citet{2014ApJ...788..121K} require $\fesc \sim 0.1$ and \citet{Wise:2008cs} obtain $\fesc \sim 0.5$ for galaxies with halo masses $M_h \lesssim 10^{8} M_\odot$ when studying dwarf galaxies near the star-formation limit.
This is verified by
\citet{Yajima:2010ij} who find $\fesc \sim 0.4$ at $3 \lesssim z \lesssim 6$ in a similar mass range.
These high escape fractions contrast with much lower values of $\fesc \lesssim 0.03 - 0.05$ from \citet{gnedin2008escape} and \citet{Ma:2015nya}.
Given the large uncertainty in the theoretical models, constraints on $\fesc$ that do not rely on specific expectations on the shape of the redshift evolution will be valuable for comparing to a broad range of models.

In this paper we revisit the problem of reconstructing the redshift evolution of both the escaped LyC photon production rate density $\ngamma(z)$ and the escape fraction $\fesc(z)$ from early galaxies.
Both parametric and non-parametric methods have been used previously to find constraints on these functions. For example, \citet{Mitra:2010sr,Mitra:2011uv,Pandolfi:2011kz,Mitra:2012av,Mitra:2015yqa} have used principle component analyses to infer both quantities with Lyman-$\alpha$, QSO, WMAP, and \emph{Planck} 2015 data; \citet{Kuhlen:2012vy} use parametric reconstruction methods for WMAP7 and $z \sim 4$ Ly$\alpha$ forest data; \citet{jensen2016machine} have used L1-regularized regression (lasso) with mock JWST/NIRSpec data; and \citet{Bouwens:2015vha} have considered parametric constraints on the time-dependent production rate of ionizing photons that escape into the IGM using \emph{Planck} 2015 data.

In light of this, our present work contains a couple of novel results.  First, we compare the constraints on $\fesc(z)$ and $\ngamma(z)$ as obtained from the latest measured values of the Thomson optical depth $\tau$ with EE low-$\ell$ polarization from~\citet{Aghanim:2016yuo} to different CMB and BAO data combinations that are in mild-to-moderate tension with each other.  The value of $\tau$ as obtained from combinations of \emph{Planck} 2015 and 2016 temperature, polarization, and lensing data and 6dFGS, SDSS Main Galaxy Sample, BOSS-LOWZ, and CMASS-DR11 BAO measurements, varies within a few standard deviations~\citep{Beutler:2011hx,Anderson:2013zyy,Ross:2014qpa,Ade:2015xua,Ade:2015zua,Aghanim:2016yuo}.  For instance,
the \emph{Planck} 2015 analysis~\citep{Ade:2015xua} obtains $\tau = 0.079 \pm 0.017$ at 68\% confidence, using the temperature and $E$-mode polarization anistropies' auto- and cross-spectra (TT, TE, EE + lowP), while a polarization-independent analysis that includes CMB lensing and BAO has a marginally lower value of $\tau = 0.067 \pm 0.016$.  However, using low multipole moments of the E-mode polarization power spectrum from the \emph{Planck} High Frequency Instrument (HFI) measures the considerably lower value of $\tau = 0.055 \pm 0.009$~\citep{Aghanim:2015xee}.  This considerably lower measurement of $\tau$ implies an average reionization redshift of $7.8 \lesssim \zreion \lesssim 8.8$, which is closer to the expectations from galaxy observations~\citep{Adam:2016hgk}.  Here, we explore the effects of the large difference in these values of $\tau$ on the LyC flux that escapes into the IGM.

Furthermore, we provide the first manifestly Bayesian, model-independent reconstruction scheme for $\ngamma(z)$ and $\fesc(z)$.
Our non-parametric redshift method is a uniform stochastic process that interpolates the functional form of $\ngamma(z)$ and $\fesc(z)$ over a high-dimensional vector in redshift space.  This gives significant freedom in the models' ability to fit the data.  Similar techniques have been used in cosmology to reconstruct the primordial power spectrum of curvature perturbations as a function of scale~\citep[\eg][]{Bridges:2005br,Bridges:2006zm,Verde:2008zza,Bridges:2008ta,Peiris:2009wp,Bird:2010mp,Vazquez:2012ux,Vazquez:2011xa,dePutter:2014hza,Aslanyan:2014mqa,Abazajian:2014tqa,Ade:2015lrj} and the redshift evolution of the dark energy equation of state~\citep{Hee:2015eba}, among others.

Finally, when calculating $\fesc (z)$ we assume two possible forms of the UV GLF that are constructed to fit HST observations of galaxies at redshifts $6 \lesssim z\lesssim 10$~\citep{Oesch:2013bha,Bouwens:2014fua}, but deviate at higher redshifts.
Our first GLF model adopts the high-redshift extrapolated values from \citet{bouwens2015most}, obtained from HST observations.  Throughout this paper we will refer to this GLF model as the B15 model.
The second model adopts
the galaxy luminosity function and $\MSF(z)$ of the \textsc{Scorch I} simulation~\citep{Trac:2015rva}, which is obtained by abundance matching to dark matter halos in high resolution N-body simulations.
We also compare a non-parametric, model independent strategy for recovering $\fesc(z)$ with a simple power-law parametric analysis.

The rest of this paper is organized as follows:
Sect.~\ref{sect:tau} describes the calculation of the Thomson optical depth.
Sect.~\ref{sect:GLF} discusses the two galaxy luminosity functions we study.
Sect.~\ref{sect:model} outlines the statistical methodology.
Sect.~\ref{sect:data} reports on the data sets that we use and the priors that we place on the parameters of interest.
Sect.~\ref{sect:results} describes the data used and the results of the Monte Carlo analysis.
Sect.~\ref{sect:concl} is a discussion and conclusion.

\section{The Thomson optical depth and escaped photon production rate}
\label{sect:tau}

Measurements of the Thomson optical depth $\tau$ indirectly constrain the ionized hydrogren fraction from the present day until the surface of last scattering.
The optical depth is calculated by integrating the proper number density of free electrons $n_e^{(p)}$ along the line of sight $l$ to the CMB:
\begin{equation}
  \tau = \int \sigma_T \, n_e^{(p)} d l,
  \label{eqn:tau1}
\end{equation}
where $\sigma_T=6.6524 \e{-25} \mathrm{cm}^2$ is the Thomson cross section.  Using the fractional volume occupied by ionized hydrogen $\QHII$, also known as the volume filling factor, the Thomson optical depth can be expressed as
\begin{equation}
  \tau = \int_{0}^{z_\mathrm{LS}} dz \frac{c \left(1+z\right)^2}{H(z)} \QHII(z) \sigma_T \nH \left( 1 + \eta \frac{Y_\mathrm{He}}{4X_\mathrm{H}} \right),
  \label{eqn:tau}
\end{equation}
where $z$ is redshift, $z_\mathrm{LS}$ is the redshift of last scattering, $c$ is the speed of light, $H(z)$ is the Hubble parameter, $X_\mathrm{H}=0.747$ is the primordial hydrogen abundance~\citep{Aver:2011bw}, $Y_\mathrm{He} \approx 1-X_\mathrm{H}$ is the primordial helium abundance,
and
\begin{equation}
  \eta \equiv
  \begin{cases}
    0 & \text{if only H ionized} \\
    1 & \text{if H ionized \& He singly ionized} \\
    2 & \text{if H ionized \& He doubly ionized}
  \end{cases} \, .
  \label{eqn:}
\end{equation}
We assume that the first reionization of helium occurs simultaneously with hydrogen reionization and that the second ionization of helium, which requires $54.4$ eV photons, occurs instantaneously at redshift $z=3.0$ and is mediated by quasar emission, although this has little effect on the calculation of $\tau$.
The comoving average number density of neutral hydrogen is
\begin{equation}
  \frac{\nH}{[1.88\e{-29} \mathrm{cm}^{-3}]} =  \left(\frac{X_\mathrm{H}}{m_\mathrm{H}}\right) \Omega_b h^2,
  \label{eqn:}
\end{equation}
where $m_\mathrm{H}$ is the hydrogen mass.

The volume filling factor of ionized hydrogen satisfies the differential equation~\citep{Madau:1998cd}
\begin{equation}
  \frac{d \QHII}{dt} = \frac{\ngamma}{\nH} - \frac{\QHII}{\trec},
  \label{eqn:}
\end{equation}
where the spatially averaged recombination time $\trec$ is given by~(\emph{e.g.},~\citet{Kuhlen:2012vy})
\begin{align}
  \frac{\trec}{\left[0.93 \, \mathrm{Gyr} \right]} \approx
  \left( \frac{\CHII}{3} \right)^{-1} \left( \frac{T_0}{2\e{4} \, \mathrm{K}} \right)^{0.7} \left(\frac{1 +z}{7} \right)^{-3},
  \label{eqn:}
\end{align}
where $T_0$ is the temperature of the IGM at mean density and $\CHII=\vev{n_\mathrm{HII}^2}/\vev{n_\mathrm{HII}}^2$ is the effective clumping factor in ionized hydrogen.  Throughout this paper we fix $T_0 = 2\e{4} \, \mathrm{K}$, which matches the expectations from typical star-forming galaxy spectra~(\emph{e.g.},~\citet{Miralda-Escudé15011994,Hui:2003hn,Trac:2008yz}).

The number of LyC photons emitted into the diffuse IGM per unit time per unit comoving volume is obtained by integrating the UV GLF over all magnitudes below the minimum star-formation magnitude $\MSF$, weighted by the amount of LyC luminosity at a given $\MUV$.  This relationship is given by
\begin{equation}
  \ngamma(z) = \fesc(z) \, \xiion \int_{-\infty}^{\MSF(z)} \phi(\MUV,z) \LUV d\MUV,
  \label{eqn:ngamma}
\end{equation}
where we have assumed that the hydrogen ionizing luminosity is linearly related to the UV luminosity
\begin{equation}
  L_\mathrm{ion} = \xiion \LUV,
  \label{eqn:}
\end{equation}
with LyC photon production efficiency $\xiion$.  The production efficiency is not precisely known, although $\xiion \approx 10^{25.2}$ is expected from galaxies with Population II stars~\citep{Schaerer:2002yr,Bruzual:2003tq}. \citet{bouwens2015using} measures $\xiion \approx 10^{25.3 \pm 0.1}$ with lower redshift $z\sim 4-5$ galaxies.
In this paper we will marginalize our results over the more conservative range $\xiion = 10^{23.5}-10^{27.5}$, as the photon production efficiency is degenerate with the overall amplitude of $\fesc$ as measured by $\tau$.  While eestricting this range would provide a tighter constraint on the typical scale of $\fesc$, it would have little effect on the recovery of any redshift evolution.

If the UV GLF is parametrized with the typical Schecter form
\begin{equation}
  \phi = \phi_* \left( \frac{\LUV}{L_*} \right)^\alpha \exp \left(-\frac{\LUV}{L_*} \right),
  \label{eqn:}
\end{equation}
then the escaped LyC photon production rate can be expressed as
\begin{equation}
  \ngamma = \left( \frac{2.5}{\log 10} \right) \xiion \, \fesc \, \phi_* \, L_* \; \Gamma \left(1+\alpha, \frac{L_\mathrm{SF}}{L_*} \right),
  \label{eqn:ngamma_int}
\end{equation}
where $\Gamma$ is the incomplete Gamma function, $L_\mathrm{SF}$ is the minimum star-formation luminosity limit, and
UV magnitudes and UV luminosities are related by the AB relation
\begin{equation}
  \frac{\LUV}{\left[\mathrm{erg\ }\mathrm{s}^{-1} \mathrm{Hz}^{-1}\right]} = \left(4.345\e{20} \right) \, 10^{-\MUV/2.5}.
  \label{eqn:}
\end{equation}
If $\alpha < -2.0$, then Eq.~\eqref{eqn:ngamma} is closely approximated by
\begin{equation}
  \ngamma = \left( \frac{2.5}{\log 10} \right) \left( \frac{ \xiion \, \fesc \, \phi_* \, L_*}{1+\alpha } \right) \left( \frac{L_\mathrm{SF}}{L_*} \right)^{1+\alpha},
  \label{eqn:ngamma_approx}
\end{equation}
where $L_\mathrm{SF}$ is the UV luminosity of galaxies at UV magnitude $\MSF$.

\section{Galaxy luminosity functions for recovering the escape fraction}
\label{sect:GLF}

In order to recover $\fesc(z)$ from the measurements of the Thomson optical depth $\tau$, we first need to specify a galaxy luminosity function that can be substituted into Eq.~\eqref{eqn:ngamma}.  \citet{Oesch:2013bha} and \citet{Bouwens:2014fua} present HST observations of $\phi$ at redshifts $z \lesssim 10$ that provide constraints over the rest frame UV magnitude range $-23 \lesssim \MUV \lesssim -16$, while \citet{Livermore:2016mbs} obtain a GLF consistent with \citet{Oesch:2013bha} and \citet{Bouwens:2014fua} up to $\MUV \lesssim 13$ at $z \gtrsim 6$ in the Hubble Frontier Field.  At higher redshift and at lower luminosities, observations do not currently provide much information and we need to rely on extrapolations from empirical models or numerical simulations to calculate $\tau$.

In this paper, we compare the reconstructed $\fesc(z)$ when using both an empirical and simulated galaxy luminosity function.
For an empirical GLF, we choose the best-fit values from \citet{bouwens2015most}, which are derived from observations reported in \citet{Bouwens:2014fua}.  This model has the Schecter parameters $M_* =-20.97 + 0.17 (z-6.0)$, $\phi_* = \left(4.5 \e{-4}\right)  10^{-0.21(z-6.0)} \mathrm{Mpc}^{-3} $, and $\alpha = -1.91 - 0.13(z-6.0) $, where the values at $z \gtrsim 8$ and $\MUV \gtrsim -15$ are based on extrapolation.  To calculate the intrinsic LyC photon production rate we use a fiducial limiting star formation magnitude of $\MSF=-10.0$, which is the commonly assumed limit (\emph{e.g.},~\citet{2012ApJ...752L...5B,2013ApJ...768...71R}), but is not directly constrained.  We allow the limiting magnitude to vary with redshift, which we parametrize by
\begin{equation}
  \MSF(z) = -10.0 + \left(\frac{d\MSF}{dz} \right) (z-6.0),
  \label{eqn:dMSF}
\end{equation}
where we restrict $-1.0 < d\MSF/dz < 0.5$.  The redshift evolution of the limiting magnitude at redshifts $z>6$ is expected to reach the ranges $-14 \lesssim \MSF \lesssim -12$ from \citet{Gnedin:2016vdz} and $\MSF \sim -12 $ from \citet{Trac:2015rva} and measurements of lensed galaxies in the Hubble Frontier Field have probed down to $\MSF = -15$ at $z=8$~\citep{Livermore:2016mbs}.  These limits qualitatively match the range we have allowed for $d\MSF/dz$.  We call this empirical GLF the B15 model.

As a test of the empirical luminosity function, we compare the B15 model to the GLF from the Scorch simulation of \citet{Trac:2015rva}, which has been calibrated to match HST observations at $z = 6$ and $\MUV \lesssim -15$ within the quoted error.  At higher redshifts and lower luminosities the luminosity function is obtained by abundance matching to a high-resolution $\Lambda$CDM N-body simulation.  This is a physically motivated cosmology model for $\phi(z)$ at high redshift, in contrast to a straightforward extrapolation from data.  The Schecter parameters for this model are expressed in Table~\ref{table:Scorch}.
\citet{Trac:2015rva} also predicts a known redshift dependence $\MSF(z)$, which does not require us to marginalize over the time evolution of the limiting magnitude.
We call this GLF the Scorch model.

\begin{figure}
\centering
\includegraphics[width=0.5\textwidth]{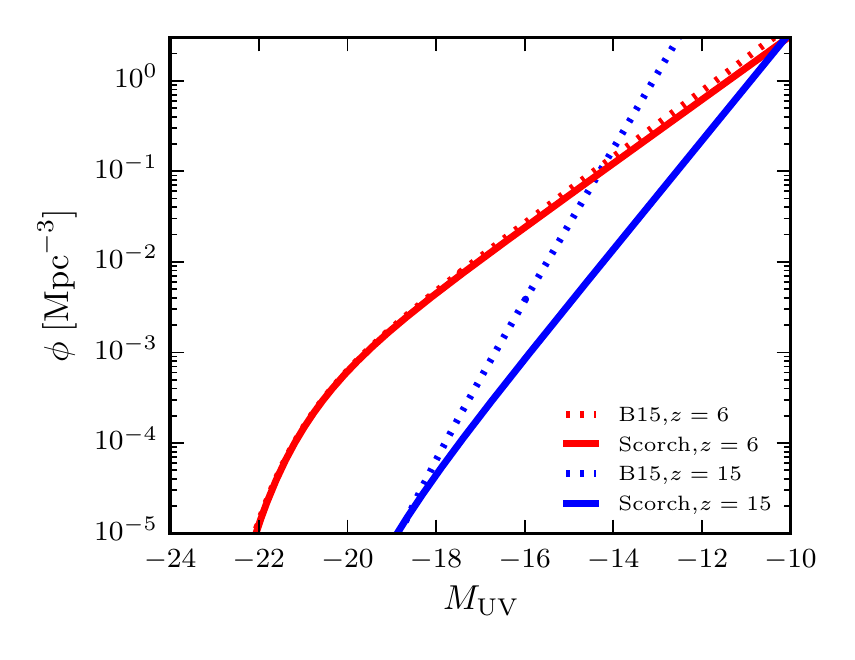}
\caption{Comparison of the galaxy luminosity function $\phi(z)$ for the Scorch model with the empirical B15 luminosity function~\citep{bouwens2015most}, as a function of UV magnitude.
}
\label{fig:phi_vs_Muv}
\end{figure}

\begin{deluxetable}{l  c c  c c  c }
  \tablewidth{\hsize}
  \tabletypesize{\footnotesize}
  \tablecaption{\label{table:Scorch} Scorch galaxy luminosity function Schecter parameters.}
  \tablecolumns{5}
  \tablehead{$z$ & $\phi_*$ & $M_*$ & $\alpha$ & $\MSF$}
  \startdata
  6   &  4.86\e{-4}  & -20.92 &  -1.88  &  -9.7 \\
  7   &  3.76\e{-4}  & -20.76 &  -1.94  & -10.0 \\
  8   &  2.67\e{-4}  & -20.61 &  -2.00  & -10.4 \\
  9   &  1.77\e{-4}  & -20.47 &  -2.07  & -10.7 \\
  10  &  1.11\e{-4}  & -20.33 &  -2.14  & -11.1 \\
  11  &  6.65\e{-5}  & -20.19 &  -2.21  & -11.4 \\
  12  &  3.84\e{-5}  & -20.06 &  -2.28  & -11.7 \\
  13  &  2.13\e{-5}  & -19.92 &  -2.35  & -12.0 \\
  14  &  1.15\e{-5}  & -19.79 &  -2.43  & -12.3 \\
  15  &  6.01\e{-6}  & -19.66 &  -2.50  & -12.6
  \enddata
  \tablecomments{The cosmological parameters are fixed at $\Omega_m = 0.3$, $h=0.7$, $\sigma_8 = 0.82$ and $\phi_*$ is in units of $\mathrm{Mpc}^{-3}$.}
\end{deluxetable}

Since the Scorch GLF is derived from a cosmology simulation, we have tested that the predictions are not sensitively dependent on the exact values of the $\Lambda$CDM parameters that are chosen.
When replicating the techniques of \citet{Trac:2015rva} for cosmological models with different values of $\Omega_m$ and $\sigma_8$,
we find that there is little variation in $\phi$, assuming that these two cosmological parameters are not allowed to vary substantially from the best-fit values on these parameters obtained by \emph{Planck} 2015 TT,TE,EE+lowP~\citep{Ade:2015xua}.  Consequently, in our analysis we will fix
the Schecter parameters of the Scorch GLF to those obtained with cosmological parameters set to $\Omega_m = 0.30$, $h=0.70$, $\sigma_8=0.82$.

Fig.~\ref{fig:phi_vs_Muv} shows the two GLFs at redshifts $z=6$ and $z=15$.  The two agree very well at $z=6$ by design, while the B15 model extrapolated to $z=15$ generically predicts more galaxies at a given $M_\mathrm{UV}$.  For fixed $\fesc$, this difference at high redshift will give earlier $\zreion$ for the B15 model than the Scorch model unless $d\MSF/dz < 0$.

\section{Modeling the evolution of the escaped photon production rate and the UV escape fraction}
\label{sect:model}

Our goal in this paper is to obtain constraints on the escaped photon production rate density $\ngamma(z)$ and the LyC escape fraction $\fesc (z)$ using parametric and non-parametric methods and compare the results when different datasets are used.
Since the redshift evolution of $\ngamma$ has been constrained by~\citet{Bouwens:2015vha} assuming a simple exponential relationship $\ngamma \sim 10^{B(z-8)}$, we will focus only on non-parametric functional forms for the escaped photon production rate.

An obvious parametric form for the escape fraction is a simple, two-parameter power-law fit
\begin{equation}
  \fesc(z) = f_8 \left( \frac{1+z}{9} \right)^\beta,
  \label{eqn:power}
\end{equation}
where $f_8$ is the escape fraction at $z=8$.  We require that the slope $\beta$ is greater than unity, in order to ensure that the escape fraction decreases with decreasing redshift.

Our non-parametric method for reconstructing $\fesc(z)$ and $\ngamma(z)$ is a uniform stochastic process.
We define a grid of $N$ evenly spaced points in redshift that belong to the range $3 \le z \le 15$.
We then uniformly sample each component of an $N$-dimensional target vector $\vec F$ identically and independently from a uniform distribution, with ranges depending on which function we are attempting to reconstruct, $\fesc$ or $\ngamma$.  We will associate $\vec F$ with the independent variable in the reconstruction at the gridded $z$-positions.
For a given $\vec F$, we can then define $\fesc(z)$ or $\ngamma(z)$ by interpolating between $\vec F$ at the gridded locations in $z$ using an $n^\mathrm{th}$-order polynomial, with any required values in the redshift ranges $z<3$ and $z>15$ fixed by the first or last value in $\vec F$, respectively.

When reconstructing the escape fraction functional form we sample the elements of the target function $\vec F$ over the range
\begin{equation}
  \vec F = \fesc \big |_{z=3,\dots,15} \sim \; \mathcal U[0,1]
  \label{eqn:}
\end{equation}
and use a cubic spline interpolation function.  We use four ``knots'' for the spline, \emph{i.e.} the dimensionality of $\vec F$ is four.
When the interpolation procedure naively exceeds the bounds $0\le \fesc(z) \le 1$, we fix $\fesc$ to the bounded value.
We further impose the constraint that $\fesc(z)$ evolves monotonically over the whole reconstruction range $3 \lesssim z \lesssim 15$, by stipulating $d \fesc/dz \ge 0$ for all $z$.

When reconstructing the escaped photon production rate density directly, we instead sample the target function in log-space as
\begin{equation}
  \vec F = \log_{10} \left( \frac{\ngamma /\nH  }{\mathrm{Gyr}^{-1}} \right) \biggr |_{z=3,\dots,15} \sim  \; \mathcal U[1,6]
  \label{eqn:}
\end{equation}
and use a linear spline.  Here, we use six knots.
We do not impose monotonicity on this functional reconstruction.

In principle, increasing the dimension of $\vec F$ by adding more internal knots to the interpolation function, or allowing the $z$-position of the knots to vary will change the shape of the functions that we are able to reconstruct.
However, we keep these hyperparameters fixed in this analysis, since the Thomson optical depth is an integrated quantity over redshift and is not sensitive to fine structure in $\fesc(z)$ or $\ngamma(z)$.
The prior probability that we use on the shape of function via this method is therefore weighted toward functions that are relatively smoothly varying in $3 <z <15$.

\section{Data and Methods}
\label{sect:data}

To obtain constraints on
$\fesc(z)$ and $\ngamma(z)$ at $z \gtrsim 6.0$ we use the measured value of $\tau$ from the combined analysis of \emph{Planck} 2015 and 2016 temperature, polarization, and lensing data and 6dFGS, SDSS Main Galaxy Sample, BOSS-LOWZ, and CMASS-DR11 BAO measurements~\citep{Beutler:2011hx,Anderson:2013zyy,Ross:2014qpa,Ade:2015xua,Ade:2015zua,Aghanim:2016yuo}.  We focus on three different measurements of $\tau$, using different combinations of these data.  First, we look at the value as predicted by the \emph{Planck} 2015 data, which is $\tau = 0.079 \pm 0.017$ based on the TT, TE, EE, + lowP data likelihood for the temperature and polarization power spectra.  We then use the lower value of $\tau = 0.067 \pm 0.016$ when BAO observations are included, but CMB polarization data is excluded.  Finally, we compare these to the latest \emph{Planck} 2016 lowE observations of $\tau = 0.055 \pm 0.009$.

We give zero likelihood to those models that have $\QHII<0.79$ at $z=5.9$, which corresponds to the $3 \sigma$ lower limit on $\QHII$ from \citet{McGreer:2014qwa} and matches the earlier constraints from~\citet{McGreer:2011dm}.
These limits are derived by quasar observations and are broadly consistent with the hydrogen reionization fraction obtained by measuring the Gunn-Peterson optical depth from~\citet{Fan:2005es}.
When constraining $\fesc(z)$ we also limit $\fesc < 0.1$ and $z <3.3$, which matches the $2\sigma$ limit from \citet{Boutsia:2011mk} and is consistent with~\citet{2007ApJ...667L.125C,Iwata:2008qx,2016arXiv160201555S}.

We do not impose any constraints on the escaped photon production rate $\ngamma$ when performing the direct non-parametric reconstruction of this quantity, \emph{e.g.}, $\ngamma/\nH = 10^{50.99 \pm 0.45}$ $\mathrm{s}^{-1}$ $\mathrm{Mpc}^{-3}$ at $z=4.75$ from \citet{Becker:2013ffa}.
However, we have checked that imposing these upper and lower limits on post-processed MCMC chains does not radically alter our non-parametric constraints on $\ngamma(z)$, particularly at higher redshifts $z \gtrsim 6$.  This contrasts with parametric reconstructions of this quantity, where the high and low redshifts values of $\ngamma$ must be simultaneously matched by a specific functional form.
For simplicity, we do exclude those scenarios where $\ngamma/\nH < \trec^{-1}$ at any point after the complete reionization $\QHII=1$, \emph{i.e.}, we do not allow any periods of possible recombination in the late universe.  This requires $\fesc > 0$ at low redshift.

We do not impose any upper bound on the reionization redshift or the \textsc{Hii} volume filling factor, \emph{e.g.}, those obtained by Ly$\alpha$ emissions or damping wings in quasar spectra~\citep[\eg][]{Schroeder:2012uy,Pentericci:2014nia,Schenker:2014tda}.  We do not include these constraints because the CMB generically predicts a higher redshift of reionization ($\zreion \sim 8-10 $) than Ly$\alpha$ ($\zreion \lesssim 6-8$).  Since these two observations are in tension, we use only the high redshift CMB and BAO datasets for upper limits derived from $\tau$.  Importantly, we do note that this conflict in the measurement of $\QHII$ is reduced with \emph{Planck} 2016 lowE data.

\begin{figure*}
\centering
\includegraphics{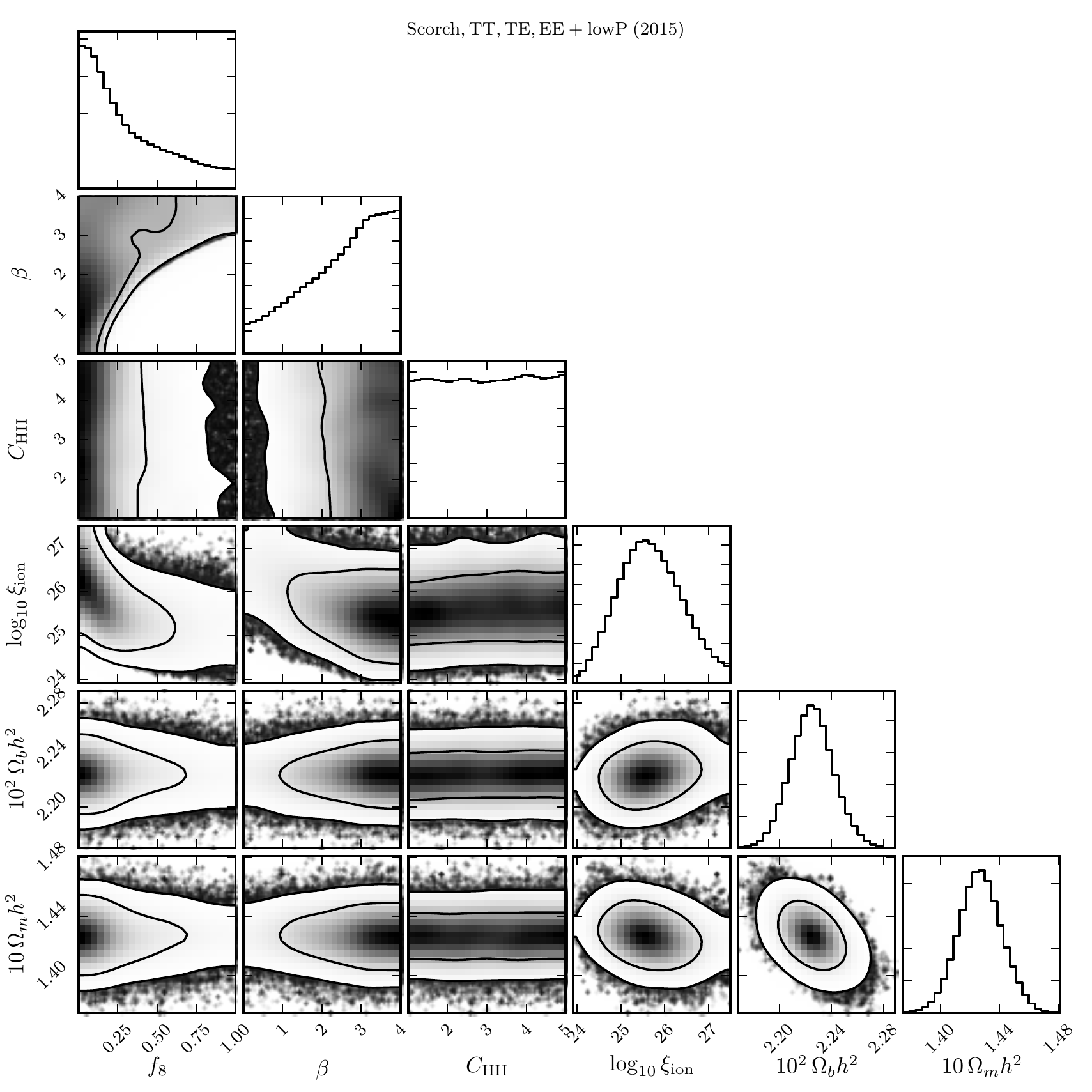}
\caption{68\% and 95\% credible regions (CRs) on the parameters of the Scorch reionization model and one-dimensional marginalized posterior probabilities, as constrained by \emph{Planck} (2015) TT, TE, EE, and lowP data.  The escape fraction $\fesc(z)$ is parametrized with the power-law model of Eq.~\eqref{eqn:power}, with amplitude $f_8$ at $z=8$ and slope $\beta$.  The effective clumping factor $\CHII$ is essentially unconstrained, while the photon production efficiency peaks near $\xiion=10^{25.5}$.  The physical baryon and matter densities ($\Omega_b h^2$ and $\Omega_m h^2$) fix the neutral hydrogen density and affect the ionization rate in the IGM, but show no degeneracy with the other parameters, as expected.}
\label{fig:power_corner}
\end{figure*}

In all cases we place prior probabilities on the parameters in the model and do a Monte Carlo search through parameter space to identify Bayesian credible regions (CRs) at the 68\% and 95\% level.
We use a flat prior on the effective clumping factor over the range $1 \le \CHII \le 5$, where the lower limit is the minimum value possible and the upper limit is set by comparison to simulations~\citep[\eg][]{Pawlik:2009ij,Kaurov:2014gra}.
However, since the LyC photon production efficiency can potentially vary over more than an order of magnitude, we use a log-flat prior on $\xiion$ in the range $23.5 \le \log_{10} \xiion \le 27.5$, with these bounds set so that it exceeds any range that has significant probability from the data-likelihoods alone.
When marginalizing the cosmological parameters, we use the same priors as the $\Lambda$CDM from \citet{Ade:2015xua}.

Our three analyses and corresponding prior probabilites for $\fesc(z)$ and $\ngamma(z)$ are
\begin{enumerate}

  \item Parametric $\fesc(z)$, using GLFs:
    for both the B15 and Scorch GLFs, we place a flat prior on the value of the escape fraction at $z=8$, $0 \le f_8 \le 1$ and a flat prior on the slope of Eq.~\eqref{eqn:power} in the range $0 \le \beta \le 4$.  Using a higher value for the slope is largely redundant, since a very steep increase in the escape fraction at $z \sim 8$ can be effectively modeled by a step function that is adequately captured by $\beta \sim 4$.

  \item Non-parametric $\fesc(z)$, using GLFs:
    for both the B15 and Scorch GLFs, we uniformly sample the value of $\fesc(z)$ between zero and unity at the ``knot'' positions $z=3,6,9,12$ according to the method of Sect.~\ref{sect:model}, and use a cubic spline to interpolate $\fesc(z)$ for intermediate values.  We require monotonicity, where $d\fesc/dz \ge 0$.

  \item Non-parametric $\ngamma(z)$, GLF-independent:
    we sample $\log_{10} \ngamma/\nH$ uniformly at the ``knot'' positions $z=3,6,9,12,15,18$, using a linear interpolation, following Sect.~\ref{sect:model}.

\end{enumerate}

\begin{figure*}
\centering
\includegraphics{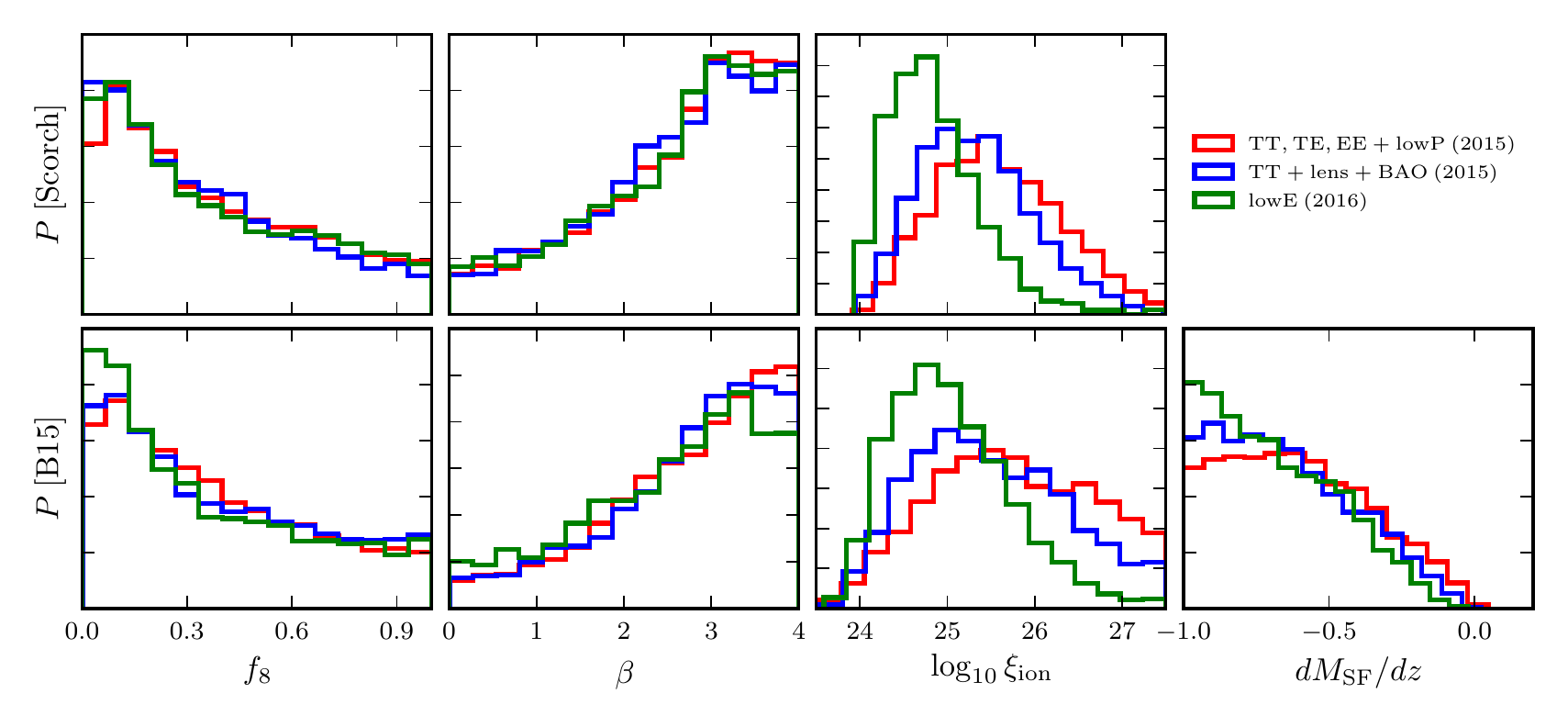}
\caption{
  Comparison of marginalized posterior probability distributions for the Scorch and B15 models using different data.
  The escape fraction $\fesc(z)$ is the power law of Eq.~\eqref{eqn:power} and the B15 limiting magnitude varies according to Eq.~\eqref{eqn:dMSF}.  We have fixed the cosmology parameters $\Omega_b h^2 = 0.022$, $\Omega_b h^2 = 0.15 $.
  The $f_8$ and $\beta$ constraints are largely consistent for both models and all data, while $\xiion$ should vary with the data in order to recover the variation in the measured Thomson optical depth.
  The empirical B15 model generically requires redshift evolution in the limiting magnitude, $d\MSF/dz <0$.
}
\label{fig:1d_hist}
\end{figure*}

To calculate the posterior probability for a set of model parameters given the data we importance sample the publicly available \emph{Planck} MCMC chains.  These were obtained using the \textsc{Camb} Boltzmann solver~\citep{Lewis:1999bs,Howlett:2012mh} and the \textsc{CosmoMC}~\citep{Lewis:2002ah} statistical sampler.
To sample the data-based, posterior probability distributions for these models, we use the publicly available code \textsc{emcee}~\citep{ForemanMackey:2012ig}, which implements an affine-invariant ensemble sampler.  This method efficiently explores the posterior probability function even in the case of strong degeneracy between parameters.

\section{Results}
\label{sect:results}

\subsection{Parametric LyC escape fraction}
\label{ssect:powerlaw}

Figure~\ref{fig:power_corner} shows histograms and the 68\% and 95\% credible regions (CRs) of the one- and two-dimensional posterior probability functions of the model parameters for the power-law parametric reconstruction of $\fesc(z)$ from Eq.~\eqref{eqn:power}, using the Scorch galaxy luminosity function and the fiducial \emph{Planck} 2015 TT, TE, EE + lowP data ($\tau = 0.079 \pm 0.017$).  The results in each panel are marginalized over all the other parameters with priors specified in Sect.~\ref{sect:data}.
As a consistency check, the contours for the cosmological parameters $\Omega_b h^2$ and $\Omega_m h^2$ recover that of the standard \emph{Planck} 2015 analysis~\citep{Ade:2015xua}.

With this functional form, the Scorch GLF requires a relatively large escape fraction at $z=8$, with $f_8 \lesssim 0.51$ ($68\%$ CR), although the one-dimensional posterior peaks at the lower value of $f_8 = 8.9\e{-2}$.
There is strong degeneracy between $f_8$ and $\beta$, since these two quantities together control the integrated fraction of intrinsic LyC photons that escape into the IGM.
There is little constraint on $\beta$
for small $f_8 \lesssim 0.25$, while larger power-law exponents are needed at larger $\fesc$ to satisfy the constraint $\fesc < 0.1$ at $z < 3$ from~\citet{2007ApJ...667L.125C,Iwata:2008qx,Boutsia:2011mk,2016arXiv160201555S}.
This results in $\beta > 2.36$ ($68\%$ CR) from the marginalized posterior probability, although there is a non-negligible probability for $\beta = 0.0$.
However, this lower limit must be interpreted with care, since it could have been artificially increased by setting a larger upper limit than $\beta < 4.0$, since the predictions for $\tau$ are approximately degenerate at the upper limit of the power-law slope's range.

The flat marginalized posterior for $\CHII$ demonstrates that our analysis is almost completely independent of the effective clumping factor, since this only weakly affects the recombination time over the prior range $1 < \CHII < 5$.
The LyC photon production efficiency is constrained in the range $\log_{10} = 25.63 \pm 0.69$ (68\% CR), despite the fact that it is degenerate with $f_8$.  The range of this constraint is larger than that obtained by \citet{Becker:2013ffa} at $z=4.5$, but we emphasize that this is obtained only from the CMB.

These conclusions also hold when using different datasets or the B15 empirical GLF.
In Fig.~\ref{fig:1d_hist} the one-dimensional marginalized posteriors for the power-law $\fesc(z)$ model are plotted for both GLFs, using \emph{Planck} 2015 TT, TE, EE + lowP ($\tau = 0.079 \pm 0.017$), \emph{Planck} 2015 TT, CMB lensing, and BAO ($\tau = 0.067 \pm 0.016$), and \emph{Planck} 2016 lowE ($\tau = 0.055 \pm 0.009$).  We marginalize over $\CHII$ in all cases, but do not plot the results as they replicate the flat posterior from Fig.~\ref{fig:power_corner}.

The constraints on $\fesc(z)$ are largely insensitive to the dataset used or the choice of GLF, although for \emph{Planck} 2016 lowE there is more probability for $f_8 <0.06$ than the other data, since this has the lowest $\tau$ value.  At 68\% confidence the inferred parameters are $f_8^\mathrm{Scorch} < 0.47,0.46,0.45$, $f_8^\mathrm{B15} = 0.47,0.47,0.39$, $\beta^\mathrm{Scorch} > 2.23,2.26,2.23$, and $\beta^\mathrm{B15} > 2.33,2.33,2.00$ for \emph{Planck} 2015 TT,TE,EE+lowP, \emph{Planck} 2015 TT,lensing,+BAO, and \emph{Planck} 2016 lowE, respectively.

\begin{figure*}
\centering
\includegraphics{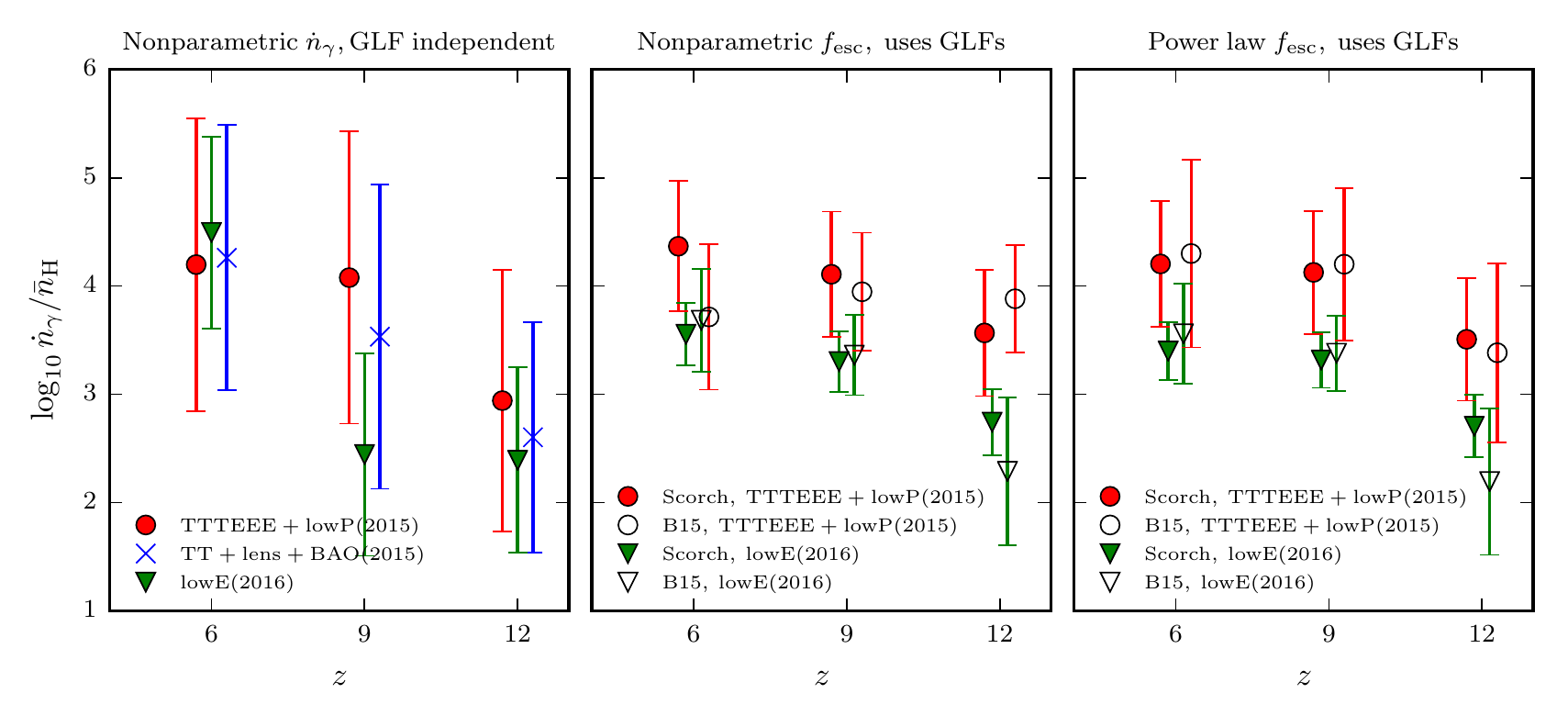}
\caption{Means and 68\% confidence intervals on the escaped LyC photon production rate density $\ngamma (z)$ at high redshift compared to the comoving neutral hydrogen density $\nH$, in units of $\mathrm{Gyr}^{-1}$.  The results are the posterior probabilities for three models: (\emph{left}) a model independent, non-parametric reconstruction of $\ngamma (z)$; (\emph{middle}) a non-parametric reconstruction of $\fesc(z)$, using the Scorch and B15 galaxy luminosity functions; and (\emph{right}) the power-law $\fesc (z)$ model (Eq.~\eqref{eqn:power}) for the same GLFs.
}
\label{fig:ngammaerror}
\end{figure*}

However, the change in the measured value of $\tau$ with the different data does alter the inferred median value of the LyC photon production efficiency $\xiion$, with similar trends between the Scorch and B15 luminosity functions.
At 68\% CR, the \emph{Planck} 2016 lowE data with the Scorch and B15 models obtain $\log_{10} \xiion^\mathrm{Scorch} = 24.82^{+0.63}_{-0.44}$ and $\log_{10} \xiion^\mathrm{B15} = 25.02^{+0.75}_{-0.59}$, respectively.  \emph{Planck} 2015 TT, TE, EE + lowP yields $\log_{10} \xiion^\mathrm{Scorch} = 25.59^{+0.78}_{-0.67}$ and $\log_{10} \xiion^\mathrm{B15} = 25.77^{+1.03}_{-0.98}$, while \emph{Planck} 2015 TT, lensing, + BAO gives $\log_{10} \xiion^\mathrm{Scorch} = 25.25^{+0.73}_{-0.59}$ and $\log_{10} \xiion^\mathrm{B15} = 25.45^{+0.99}_{-0.82}$.
Despite the allowed freedom in the overall amplitude of $\fesc$, the lowered value of $\tau$ in \emph{Planck} 2016 lowE is better matched by a reduction in the LyC photon production efficiency than in a lower $f_8$.  We note that a tighter prior on the range of $\xiion$, \emph{e.g.}, by using the conclusions of \citet{Becker:2013ffa}, would correspond to a greater adjustment in the $(f_8, \beta)$ plane when adopting the \emph{Planck} 2016 low multipole E-mode polarization data from HFI as compared to the \emph{Planck} 2015 temperature and polarization data than what is shown in Fig.~\ref{fig:1d_hist}.

For all datasets, the B15 model requires substantial redshift evolution in the limiting UV magnitude $\MSF(z)$ in order to adequately recover the measured Thomson optical depth.  The right panel of Fig.~\ref{fig:1d_hist} shows that there is little probability for $d \MSF/dz \ge 0$, with the 68\% CRs requiring $d\MSF/dz < -0.50,-0.54,-0.59$ for \emph{Planck} 2015 TT,TE,EE+lowP, \emph{Planck} 2015 TT,lensing+BAO, and \emph{Planck} 2016 lowE, respectively.  Again, the exact value of this upper limit on $d\MSF/dz$ is sensitive to the lower bound we place on this parameter in the prior, which we currently set to $d\MSF/dz > -1$.

However, if we force the B15 model to have no redshift evolution in the limiting magnitude, \emph{i.e.}, $d\MSF/dz = 0$ and $\MSF = -10.0$, then the constraints on the escape fraction change significantly.
Without a time varying limiting magnitude, the B15 GLF predicts substantially more intrinsic LyC photon production at $z \gtrsim 6$ than the Scorch model.  The B15 model with $\MSF=-10.0$ therefore needs a lower $f_8$ to produce the same $\tau$ as the Scorch GLF.  At 68\% confidence the constraints are $f_8^\mathrm{B15} < 0.14$ and $f_8^\mathrm{B15} <0.076$ for \emph{Planck} 2015 TT,TE,EE+lowP and 2016 lowE.
Since we require $\fesc<0.1$ for $z<3.0$, this also translates into a reduced probability that $\fesc$ must be evolving with redshift for the B15 model with $\MSF=-10.0$ compared to Fig.~\ref{fig:1d_hist}, $\beta < 1.27$ at 68\% for \emph{Planck} 2015 TT,TE,EE+lowP and $\beta < 0.33$ at 68\% for \emph{Planck} 2016 lowE.

\subsection{Non-parametric reconstruction}
\label{ssect:nonparam}

Fig.~\ref{fig:ngammaerror} compares the non-parametric reconstruction of $\ngamma(z)$ to the derived value of $\ngamma(z)$ obtained when using the Scorch and B15 GLFs with $\fesc(z)$ modeled both non-parametrically and with the power-law functional form.  The error bars in all three panels are the $1 \sigma$ ranges for the posterior probability on the escaped photon production rate $P(\log_{10} \ngamma (z) \| D)$.  However, note that the reconstructed curves $\ngamma(z)$ will not fall within all of the plotted error bars at different redshifts at $1 \sigma$ confidence.  In all cases we marginalize over $\CHII$ and, when calculating $\ngamma$ from the two GLFs ($\fesc(z)$ reconstructions), we marginalize $\xiion$ with the log-flat prior and $d\MSF/dz$ with a uniform prior for the B15 model.   We compare all three tension datasets: \emph{Planck} 2015 TT,TE,EE,+lowP ($\tau = 0.079 \pm 0.017$), \emph{Planck} 2016 lowE ($\tau = 0.055 \pm 0.009$), and \emph{Planck} 2015 TT,lensing,+BAO ($\tau = 0.067 \pm 0.016$).

\begin{figure*}
\centering
\includegraphics{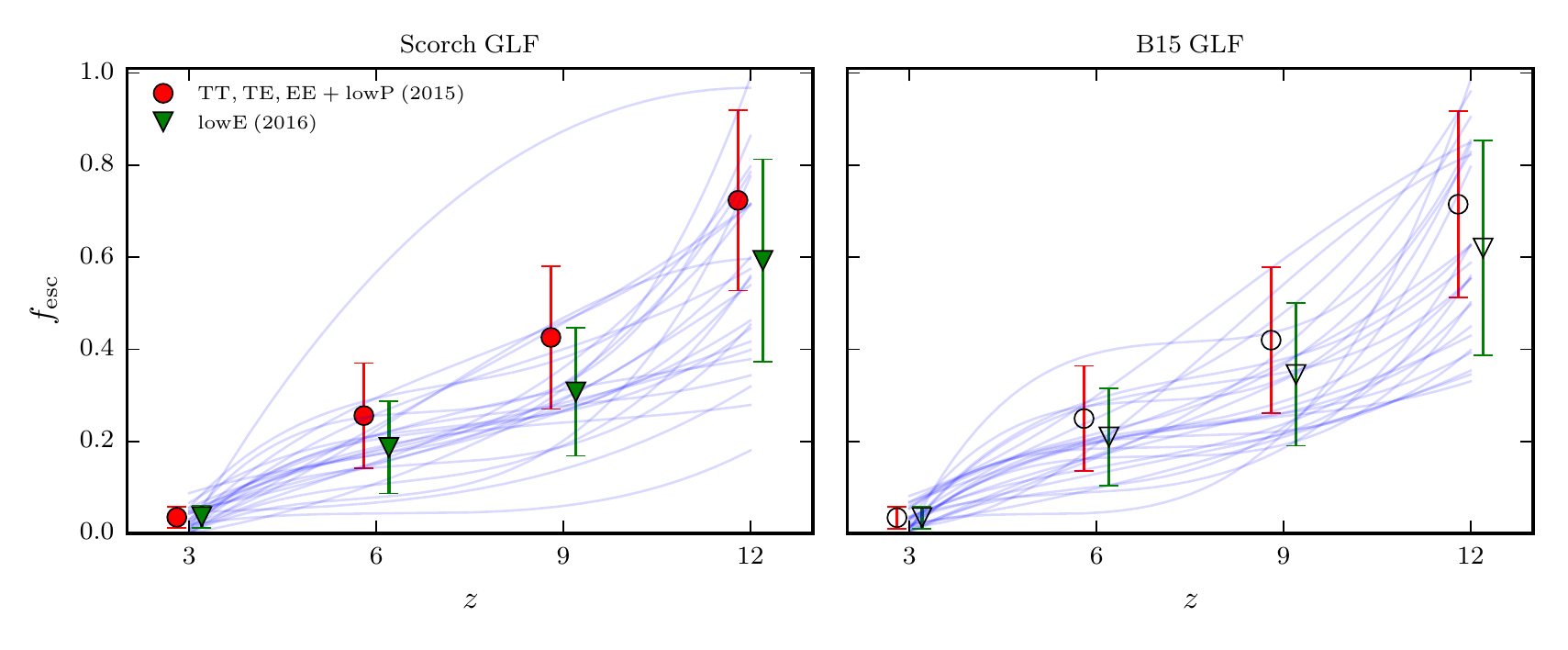}
\caption{Non-parametric reconstruction of $\fesc(z)$ when using (\emph{left column}) the Scorch GLF, (\emph{right column}) the B15 GLF with \emph{Planck} 2015 and 2016 lowE data. The dots and error bars are the mean and 68\% confidence interval of the marginalized posterior distribution $P(\fesc(z) \| D)$ at a given redshift.  The blue curves are 25 randomly chosen realizations from the posterior on the escape fraction conditioned on \emph{Planck} 2016 lowE, demonstrating the possible functional forms of the reconstructed $\fesc(z)$.  We require that $\fesc (z)$ be increasing monotonically with increasing $z$.
}
\label{fig:polinterror}
\end{figure*}

The non-parametric reconstruction of $\ngamma(z)$ is our most general result, since it does not require a pre-specified functional form, a detailed galaxy luminosity function $\phi(z)$, or a GLF limiting magnitude $\MSF(z)$.  The error bars are larger for the non-parametric $\ngamma(z)$ than for the results that utilize a GLF, since the GLF imposes a relatively tight shape on the intrinsic photon production rate.

In the non-parametric case, the lower mean value of $\tau$ with \emph{Planck} 2016 lowE results in a reduced mean value of the escaped photon production rate density at $z=9$, $\log_{10} \ngamma = 2.44 \pm 0.94$, compared to $\log_{10} \ngamma = 4.16 \pm 1.36$ (\emph{Planck} 2015 TT,TE,EE,+lowP) and $\log_{10} \ngamma = 3.54 \pm 1.39$ (\emph{Planck} 2015 TT,lensing,+BAO).
However, the recovered values are only discrepant at approximately $1 \sigma$.

At redshifts $z \gtrsim 9$, $\ngamma$ must be relatively low to stop a very early reionization epoch.
At low redshift ($z<6$) we have almost no constraining power from the CMB or BAO as the IGM has already been reionized by $z \sim 6$ and there is no further contribution to $\tau$ from $\ngamma$ at lower redshift.  The posterior probabilities on $\ngamma$ in the range $3 \lesssim z \lesssim 6$ replicate the prior probability distribution, without any additional constraints.
Importantly, this contrasts with the parametric approach in~\citet{Bouwens:2015vha} that requires the extrapolated value of $\ngamma$ at $z=4.5$ to match the empirical constraints of~\citet{Becker:2013ffa}.  Our methodology does not rely on extrapolation of $\ngamma$ in this redshift range.

When using a known GLF, the error bars on the derived value of $\ngamma(z)$ are smaller for all $z$, since a GLF imposes a definite shape on the intrinsic photon production rate.  However, the overall amplitude of $\ngamma$ is consistent between the non-parametric reconstruction and the ones that assume a GLF.
For the non-parametric $\fesc(z)$ (assuming the GLFs), we see substantial agreement between the two GLFs when constraining $\ngamma(z)$ with the \emph{Planck} 2015 and \emph{Planck} 2016 results, with the means of the posterior distributions deviating at less than $1\sigma$.
Comparing the two datasets, we see that \emph{Planck} 2016 lowE prefers a lower $\ngamma$ at all redshifts than \emph{Planck} 2015 TT,TE,EE+lowP due to its substantially lower inferred $\tau$ value.
Despite the significant freedom in the shape of $\fesc(z)$ and the overall amplitude of the escaped LyC flux as set by $\vev{\fesc} \xiion$, the inferred value of $\ngamma$ does not vary beyond about a factor of 10, demonstrating consistency.
The results are qualitatively similar when using the power-law functional form for $\fesc(z)$.
In general, the errors bars tend to be similar than in the non-parametric case, even though there is less freedom in the allowed functional form of the escape fraction, implying that a power-law fit to $\fesc$ captures much of the functional flexibility required by the data.

Fig.~\ref{fig:polinterror} shows the shape of the reconstructed $\fesc(z)$ using the non-parametric techniques described in Sect.~\ref{sect:model}.  Again, we marginalize over $\CHII$, $\xiion$, and $d\MSF/dz$ (for the B15 model).
The mean value of the escape fraction increases with increasing redshift due to our requirement of monotonicity.
The recovered functions again demonstrate the strong consistency between the results obtained using the two GLFs.
There is substantial variability in the allowed $\fesc(z)$, with no obvious difference in the shapes that are allowed by Scorch compared to B15.
Similar to Fig.~\ref{fig:ngammaerror}, \emph{Planck} 2016 lowE predicts a lower $\fesc$ than \emph{Planck} 2015 TT,TE,EE+lowP at all redshifts, as expected.  However, the difference is relatively minor given the large difference in the measured values of the Thomson optical depth for these datasets.

For \emph{Planck} 2016 lowE, the $1\sigma$ credible regions on the escape fraction are
$\fesc^{(6)} = 0.19 \pm 0.13$, $\fesc^{(9)} = 0.30 \pm 0.20$, and $\fesc^{(12)} = 0.49 \pm 0.30$
with the Scorch GLF and
$\fesc^{(6)} = 0.15 \pm 0.11$, $\fesc^{(9)} = 0.23 \pm 0.17$, and $\fesc^{(12)} = 0.40 \pm 0.29$
with the B15 GLF.
For \emph{Planck} 2015 TT,TE,EE,+lowP, the results are $\fesc^{(6)} = 0.23 \pm 0.11$, $\fesc^{(9)} = 0.37 \pm 0.16$, and $\fesc^{(12)} = 0.62 \pm 0.23$ with the Scorch GLF and $\fesc^{(6)} = 0.22 \pm 0.10$, $\fesc^{(9)} = 0.35 \pm 0.15$, and $\fesc^{(12)} = 0.57 \pm 0.23$ with the B15 GLF.  At $z=3$, $\fesc$ is distributed uniformly in the range $0 < \fesc < 0.1$, with the upper bound set by the prior from~\citet{Boutsia:2011mk}.
For \emph{Planck} 2016 lowE, the constraints on the LyC photon production efficiency are
$\log_{10}  \xiion^\mathrm{Scorch} = 24.90_{-0.43}^{+0.55}$ and
$\log_{10}  \xiion^\mathrm{B15} = 25.11_{-0.51}^{+0.67}$.
For \emph{Planck} 2015 TT,TE,EE+lowP, the constraints on the LyC photon production efficiency are
$\log_{10}  \xiion^\mathrm{Scorch} = 25.51_{-0.61}^{+0.65}$ and
$\log_{10}  \xiion^\mathrm{B15} = 24.96_{-0.61}^{+0.77}$.

As with the parametric escape fraction results, if we do not allow the limiting magnitude $\MSF$ to vary with redshift for the B15 model, then the results change dramatically.  In this case, the B15 model predicts more LyC flux at higher redshift than Scorch, which requires a lower escape fraction at all $z$.  For $d\MSF/dz=0$ with the B15 GLF, the constraints are at the much lower values: $\fesc^{(6)} = 0.26 \pm 0.12$, $\fesc^{(9)} = 0.40 \pm 0.17$, and $\fesc^{(12)} = 0.59 \pm 0.23$ (\emph{Planck} 2015 TT,TE,EE+lowP) and $\fesc^{(6)} = 0.11 \pm 0.07$, $\fesc^{(9)} = 0.14 \pm 0.09$, and $\fesc^{(12)} = 0.18 \pm 0.13$ (\emph{Planck} 2016 lowE).  We take this as evidence that the limiting magnitude varies with redshift in order to consistently match the observed value of the Thomson optical depth, unless the prior on $\fesc$ is significantly altered to prefer low values $\fesc \lesssim 0.1-0.2$.

\begin{deluxetable}{l  c c  c c  c }
  \tablewidth{\hsize}
  \tabletypesize{\footnotesize}
  \tablecaption{\label{table:bestfit} Maximum likelihood reconstructions with (\emph{top}) \emph{Planck} 2015 TT,TE,EE+lowP and (\emph{bottom}) \emph{Planck} 2016 lowE.}
  \tablecolumns{4}
  \tablehead{$z$ & $\ngamma^\mathrm{Scorch}$ & $\ngamma^\mathrm{B15}$ & $\ngamma^\mathrm{NP}$}
\startdata
  6   &  2.75\e{54} &  1.16\e{54} & 1.12\e{55} \\
  7   &  2.74\e{54} &  1.58\e{54} & 7.07\e{54} \\
  8   &  2.31\e{54} &  1.80\e{54} & 4.47\e{54} \\
  9   &  1.87\e{54} &  1.74\e{54} & 2.82\e{54} \\
  10  &  1.34\e{54} &  1.43\e{54} & 4.28\e{53} \\
  11  &  8.64\e{53} &  1.01\e{54} & 6.49\e{52} \\
  12  &  5.20\e{53} &  6.08\e{53} & 9.83\e{51} \\
  13  &  2.35\e{53} &  2.78\e{53} & 7.63\e{52} \\
  14  &  1.09\e{53} &  1.11\e{53} & 5.93\e{53} \\
  15  &  4.40\e{52} &  3.89\e{52} & 4.60\e{54} \\
  \hline
  \hline
  \\
   6  &  4.18\e{53} &  3.03\e{53} & 6.98\e{55} \\
   7  &  3.85\e{53} &  3.20\e{53} & 4.99\e{54} \\
   8  &  3.17\e{53} &  3.03\e{53} & 3.57\e{53} \\
   9  &  2.59\e{53} &  2.64\e{53} & 2.55\e{52} \\
  10  &  1.91\e{53} &  2.12\e{53} & 2.64\e{52} \\
  11  &  1.29\e{53} &  1.57\e{53} & 2.74\e{52} \\
  12  &  8.19\e{52} &  1.06\e{53} & 2.83\e{52} \\
  13  &  3.71\e{52} &  4.93\e{52} & 4.21\e{52} \\
  14  &  1.72\e{52} &  2.02\e{52} & 6.26\e{52} \\
  15  &  6.93\e{51} &  7.22\e{51} & 9.30\e{52}
  \enddata
  \tablecomments{The non-parametric LyC escape fraction $\fesc(z)$ for the Scorch and empirical B15 galaxy luminosity functions and their derived escaped LyC photon production rate density compared to the direct non-parametric (NP) reconstruction of $\ngamma$.  The units for $\ngamma$ are photons/sec/$\mathrm{Mpc}^3$.
  }
\end{deluxetable}

For convenience Table~\ref{table:bestfit} reports the best-fit values of all analyses with the latest \emph{Planck} 2016 lowE data compared with \emph{Planck} 2015 TT,TE,EE+lowP.  We give tabulated functions for $\ngamma(z)$ with and without the assumptions of the GLFs, which can be interpolated for future use.
For \emph{Planck} 2016 lowE, the maximum likelihood values for the power law model are:
$f_8 = 0.18$ and $\beta = 1.27$ (Scorch);
$f_8 = 0.08$ and $\beta = 1.39$ (B15).
For \emph{Planck} 2015 TT,TE,EE+lowP, the maximum likelihood values for the power law model are:
$f_8 = 0.18$ and $\beta = 2.55$ (Scorch);
$f_8 = 0.08$ and $\beta = 1.03$ (B15).

With the \emph{Planck} 2016 lowE data the best-fit auxiliary parameters are:
$\xiion = 10^{24.82}$ (Scorch GLF, power-law $\fesc$);
$\xiion = 10^{22.02}$ (Scorch GLF, non-parametric $\fesc$);
$\xiion = 10^{25.11}$ and $d\MSF/dz = -0.56$ (B15 GLF, power-law $\fesc$);
and
$\xiion = 10^{25.16}$ and $d\MSF/dz = -0.40$ (B15 GLF, non-parametric $\fesc$).
With the \emph{Planck} 2015 TT,TE,EE+lowP data the best-fit auxiliary parameters are:
$\xiion = 10^{25.67}$ (Scorch GLF, power-law $\fesc$);
$\xiion = 10^{25.34}$ (Scorch GLF, non-parametric $\fesc$);
$\xiion = 10^{25.00}$ and $d\MSF/dz = -0.17$ (B15 GLF, power-law $\fesc$);
and
$\xiion = 10^{25.26}$ and $d\MSF/dz = -0.41$ (B15 GLF, non-parametric $\fesc$).
For all reported best-fit values we use $\Omega_b h^2 = 0.0225$, $\Omega_c h^2 = 0.0142$, and $\CHII=3.0$.

Importantly, due to the high dimensionality of the likelihood surface and the degeneracy between different functions $\fesc(z)$ and $\ngamma(z)$ in predicting $\tau$,
there exist a number of different model configurations that are able to obtain high-likelihood fits to the data.
Consequently, there may be a family of model parameters that all yield good predictions for $\tau$, but have qualitatively different features in the escape fraction or photon production rate.
The best-fit values listed here can only be interpreted as being \emph{representative} of those functions that provide the best fit to the data.

\begin{figure*}
\centering
\includegraphics{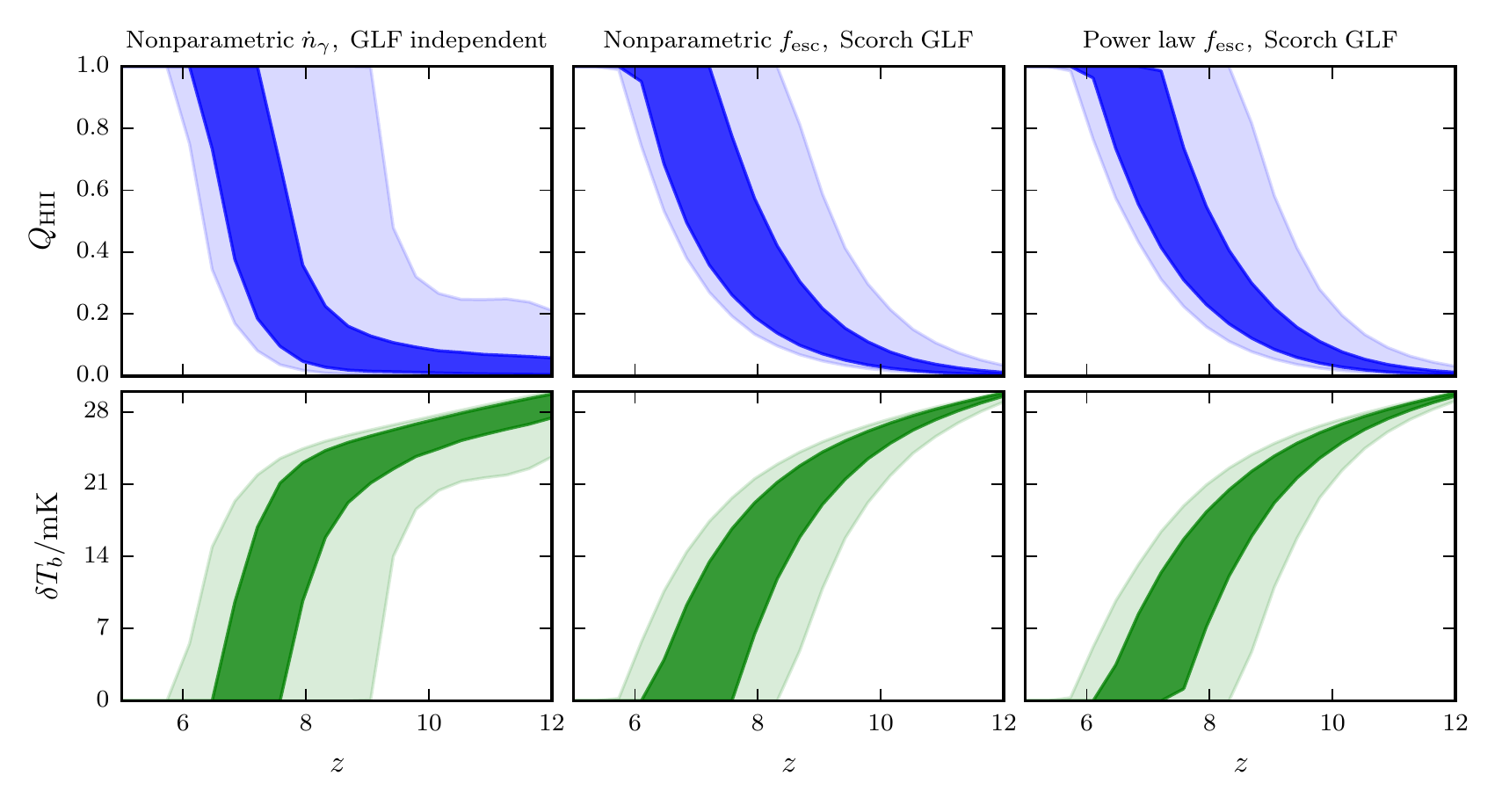}
\caption{The volume filling fraction of HII $\QHII$ as a function of redshift and the corresponding global 21cm signal $\delta T_b$, using the \emph{Planck} 2016 lowE measurement of $\tau$ and the Scorch galaxy luminosity function.  (\emph{Left column}) The direct non-parametric reconstruction of the escaped photon production rate density $\ngamma(z)$ predicts a moderately quicker reionization epoch, compared to (\emph{middle column}) the non-parametric $\fesc(z)$, and (\emph{right column}) the power-law $\fesc(z)$, which are broadly similar.
}
\label{fig:Q_and_21cm}
\end{figure*}

\subsection{HII volume filling factor and global 21cm}
\label{ssect:21cm}

The global, spatially averaged difference in intensity of the CMB as it passes through intervening neutral hydrogen is~\citep[\eg][]{Madau:1996cs,Ciardi:2003hg,Harker:2010ht}
\begin{equation}
  \frac{\delta T_b}{[25.8 \mathrm{\ mK}]} = x_\mathrm{HI} \left( \frac{\Omega_b h^2}{0.022} \right) \sqrt{ \left( \frac{0.15}{\Omega_m h^2}\right) \left( \frac{1+z}{10}\right)}
  \label{eqn:}
\end{equation}
where $T_S$ and $T_\mathrm{CMB}$ are the spin and CMB temperatures and $x_\mathrm{HI} = 1-\QHII$ is the neutral hydrogren fraction.  We assume $T_S \gg T_\mathrm{CMB}$, which is only valid for $\QHII \gtrsim 0.25$~\citep{Santos:2007dn}, but allows us to study the expected signal qualitatively at the beginning of reionization.  Measuring the global 21cm signal as a function of frequency provides information on the neutral IGM and the evolution of early sources that could potentially ionize it.  Experiments such as \textsc{Edges}~\citep{0004-637X-676-1-1} and SCI-HI~\citep{Voytek:2013nua} are targeting this signal, with \citet{2010Natur.468..796B} providing the first lower bound on the duration of reionization $\Delta z > 0.06$ by 21cm observations.
Future global 21cm signals can be used to constrain $\QHII$ directly and therefore reduce uncertainties on the inferred values of $\fesc(z)$ and $\ngamma(z)$.

Fig.~\ref{fig:Q_and_21cm} shows the volume filling factor of ionized hydrogen and the global 21cm signal from reionization driven by galaxies with the Scorch GLF and constrained by the \emph{Planck} 2016 lowE data.
For the non-parametric $\ngamma(z)$ reconstruction, the reionization epoch is relatively quicker than it is when the Scorch GLF is assumed.  This is because of the lowered value of $\ngamma$ at $z=9$ in the left column of Fig.~\ref{fig:ngammaerror} as compared to the GLF-dependent results in the middle and right columns.  Since the photon production rate can be suppressed at high redshifts if it is given enough freedom in its functional form, then it can essentially turn on a quick period of reionization in order to better fit the value of $\tau = 0.055 \pm 0.009$.  The GLF-dependent results again show a broadly consistent pattern with or without the assumption of a power-law $\fesc(z)$.  Since the 21cm signal is proportional to $\QHII$, the dependence on $z$ is qualitatively similar between $\QHII$ and $\delta T_b$

\section{Conclusion}
\label{sect:concl}

The large difference in $\tau$ as reported by \emph{Planck} 2015~\citep{Ade:2015xua} and \emph{Planck} 2016~\citep{Aghanim:2016yuo} has interesting consequences for the inferred redshift evolution of reionization parameters.
For both tested galaxy luminosity functions, we find that these require generic redshift evolution in $\fesc(z)$ even when marginalizing over the intrinsic LyC photon production efficiency $\xiion$.  However, the overall scale of $\fesc$ does not tend to be substantially alterred when constrained with different choices of data, with variations in the Thomson optical depth being better fit by changes in $\xiion$.

Our parametric and non-parametric reconstructions of $\fesc(z)$ show consistency for the derived value of the escaped LyC photon production rate density $\ngamma(z)$, which is the easiest quantity to constrain via $\tau$.
We find that a non-parametric reconstruction of the LyC photon production rate density $\ngamma(z)$ is broadly consistent with the derived value of $\ngamma(z)$ when using both parametric and non-parametric functional forms for the escape fraction $\fesc(z)$ and the Scorch and B15 GLFs.  However, the model-independent $\ngamma(z)$ predicts a relatively lower value at $z=9$ than the GLF-dependent models when using the latest \emph{Planck} 2016 lowE data, which makes the period of reionization sharper for this dataset.  The non-parametric methodology has a broader range of allowed functions $\ngamma(z)$ than when this parameter is reconstructed with parametric techniques as in~\citet{Bouwens:2015vha}, where data at $z \lesssim 3$~\citep{Becker:2013ffa} can be used to constrain the high-$z$ LyC photon production rate by extrapolation.

When using the empirically extrapolated GLF from~\citet{bouwens2015most}, we find that this model results in a generic overproduction of LyC photons compared to the Scorch model unless the limiting magnitude of the GLF decreases with increasing redshift.  Although low values of $\fesc \lesssim 0.05$ would match the $\tau\sim 0.05-0.08$ observations, it is a better fit to the data to have a time-varying limiting magnitude for the UV GLF.  In particular, since we require $d\fesc/dz \ge 0$, time-variation in the escape fraction cannot compensate for excess LyC photons at $z \sim 9-12$ compared to $z \sim 6$.  We find that $d\MSF/dz \lesssim -0.05$ at $68\%$ confidence, which is largely independent of the dataset chosen.
Therefore, we can with high confidence say that we need to incorporate redshift evolution into the limiting magnitude $\MSF$ for the empirical GLF.

When assuming a GLF but modeling $\fesc(z)$ non-parametrically, we see significant redshift evolution for all datasets, with marginally lower values of $\fesc$ when fitting to \emph{Planck} 2016 lowE than to \emph{Planck} 2015 TT,TE,EE+lowP.
When using a power-law $\fesc(z)$, the power-law exponent is similarly large $\beta \gtrsim 2.0$, demonstrating that the LyC escape fraction must evolve with time to provide the best fit to the data for the luminosity functions we have tested.

With all models for $\fesc(z)$ or $\ngamma(z)$ we find that the variation in the measured value of $\tau$ from the three tension datasets is better fit by altering the LyC photon production efficiency $\xiion$ than by changing the escape fraction, since the latter can vary only over an order of magnitude.  However, this conclusion is largely sensitive to the choice of a log-prior for $\xiion$ compared to a uniform prior for $\fesc(z)$.

By comparing non-parametric and parametric reconstructions the galactic physics of reionization we are able to more easily understand the roles that our assumptions on the galaxy luminosity function, the UV luminosity to LyC luminosity conversion rate, the escape fraction, and the limiting magnitude play in predicting $\tau$.  We advocate for this hybrid approach to studying these problems.

\acknowledgments

We thank Nick Battaglia, Rychard Bouwens, Shirley Ho, and Rachel Mandelbaum for helpful discussions.
LCP and HT are supported in part by the DOE DE-SC0011114 grant.
HT is also supported in part by NASA ATP-NNX14AB57G and NSF AST-1312991.

\bibliographystyle{yahapj}
\bibliography{references}

\end{document}